\documentclass[twocolumn,showpacs,showkeys,amsmath,amssymb,preprintnumbers,nofootinbib,floatfix,prl]{revtex4}

\usepackage{graphicx}
\usepackage{bm}
\usepackage{mathrsfs}
\usepackage{amsmath}
\usepackage{verbatim}

\begin{document}

\title{Superfluid turbulence from quantum Kelvin wave to classical Kolmogorov cascades}

\author{Jeffrey Yepez${}^{1}$\thanks{To whom correspondence should be addressed.}, George Vahala${}^{2}$, Linda Vahala${}^{3}$, Min Soe${}^{4}$}


\address{${}^{1}$Air Force Research Laboratory, Hanscom Air Force Base, Massachusetts  01731 \\
${}^{2}$Department of Physics, William \& Mary, Williamsburg, Virginia 23185\\
${}^{3}$Department of Electrical \& Computer Engineering, Old Dominion University, Norfolk, VA 23529\\
${}^{4}$Department of Mathematics, Rogers State University, Claremore, OK 23529}

\begin{abstract}
A novel unitary quantum lattice gas algorithm is used to simulate quantum turbulence of a BEC described by the Gross-Pitaevskii equation on grids up to $5760^3$.  For the first time, an accurate power law scaling for the quantum Kelvin wave cascade is determined: $k^{-3}$.  The incompressible kinetic energy spectrum exhibits very distinct power law spectra in 3 ranges of $k$-space: a classical Kolmogorov $k^{-\frac{5}{3}}$ spectrum at scales much greater than the individual quantum vortex cores, and a quantum Kelvin wave cascade spectrum $k^{-3}$ on scales of order the vortex cores.  In the semiclassical regime between these two spectra there is a pronounced steeper spectral decay, with non-universal exponent.  The Kelvin $k^{-3}$ spectrum is very robust, even on small grids, while the Kolmogorov $k^{-\frac{5}{3}}$ spectrum becomes more and more apparent as the grids increase from $2048^3$ grids to $5760^3$.
 
\end{abstract}

\pacs{47.37.+q,67.25.dk,03.75.Kk,03.75.Lm,03.67.Ac}

\keywords{quantum turbulence, BEC superfluid, Gross-Pitaevskii eq., vortex solitons, quantum lattice gas, Kelvin waves}

\maketitle

{\it Introduction.}--- Superfluid turbulence is an intriguing subject both in its own right for studying quantum turbulence in liquid Helium, Bose-Einstein condensates (BEC), and neutron stars, for example, but also for comparing to classical fluid turbulence, which itself is one of the grand challenge problems of the millenium \cite{kobayashi:045603,Barenghi20082195}.
There is a broadly acknowledged need for high resolution quantum turbulence simulations--in this Letter we strive to meet this with a unitary simulation of the Gross-Pitaevskii (GP) equation on large spatial grids up to $5760^3$.  

Fundamental to superfluid turbulence is the quantum vortex: a topological singularity with the superfluid density exactly zero at the vortex core \cite{Donnelly_1991}.
 In the simplest case, all the quantum vortices are discrete, have the same strength ({\it i.e.} same quantized circulation in multiples of $2\pi$), and the flow is inviscid.  This stands in sharp constrast to classical incompressible fluid turbulence where the concept of a vortex is imprecise, the vortices are not discrete with arbitrary size and strength, and where viscosity plays an essential role.  Moreover, in classical turbulence there are two strongly competing effects:  sweeping of small scale vortices by large scale vortices, and straining of vortices by vortices of similar scales.   Building on the idea of Richardson's local cascade of energy from large to smaller and smaller vortices till viscosity dissipates the smallest ones into heat \cite{PRSL_Richardson_1926},  Kolmogorov \cite{Kolmogorov-1941} assumed there is an inertial energy spectrum that depends only on the energy input and wave number.  
Assuming the energy transfer and the interacting scales are purely local and sweeping is not important, Kolmogorov derived the inertial energy spectrum for classical incompressible turbulence: ${E(k)=C_K \,\mathcal E^{\frac{2}{3}}}\, k^{-\frac{5}{3}}$, for some constant $C_K$.  

\begin{figure}
\includegraphics[width=3.35 in]{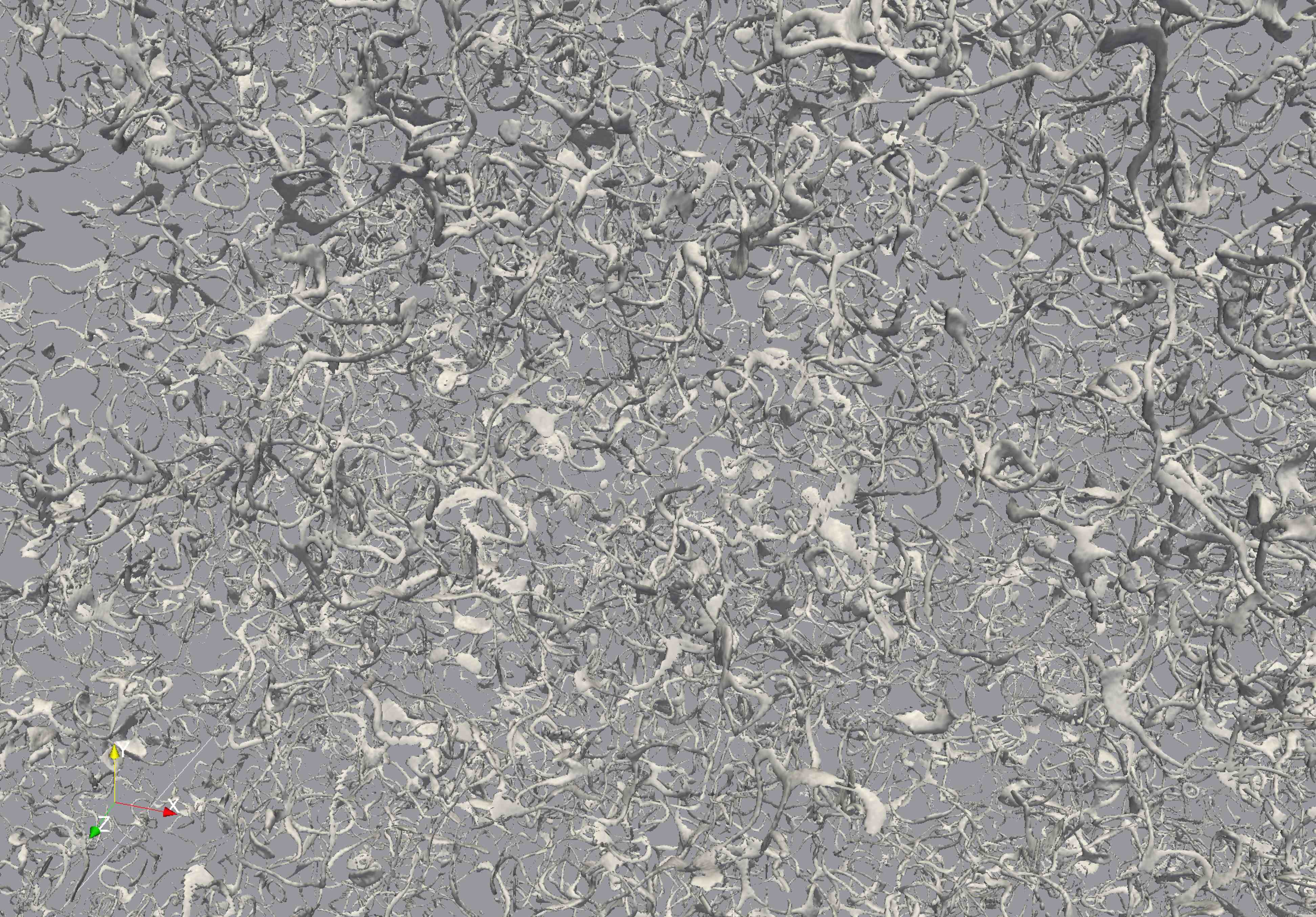}
\caption{\label{dens2_50k}\footnotesize 
Quantum turbulence (zoom-in online to view tangles).  These are vortex core isosurfaces from our quantum lattice simulation on $1024^{3}$ grid at t = 50K time step, with parameters $a=0.03$ and $g=1.004$ as defined in (\ref{Pade_approximant}).  The initial conditions are the same as for the $5760^3$-run for a repulsive nonlinear interaction.}
\end{figure}

{\it Phenomenology of quantum turbulence.}---Quantum turbulence is envisaged to arise from dense quantum vortex tangles \cite{Feynman.1955} and this is borne out by numerical simulation, for example as shown in Fig.~\ref{dens2_50k}.
 Since the flow outside a quantum vortex core is simple potential flow, it is thought that for scales $\ggg \xi$ (the radius of a quantum vortex core) the discrete nature of the quantum vortices is lost and the superfluid density is approximately constant.  Since sweeping is now much less of an issue, large scale quantum turbulence could resemble incompressible classical turbulence with Kolmogorov energy cascade 
\(
{E(k)}\approx k^{-\frac{5}{3}}
\),
for $k\ll \xi^{-1}$.

For length scales on the order of the vortex core one needs to consider the effects of vortex reconnection---a reconnection that occurs in superfluids without the need of any dissipative mechanism, unlike classical vortex reconnection processes.  During the quantum vortex-vortex reconnection/collision, the vortex lines are sharply distorted and emit large amplitude Kelvin waves (large relative to the wavelength).  The Kelvin waves now interact with each other to generate Kelvin waves of smaller and still smaller wavelength.  This Kelvin wave energy cascade continues until one reaches such short scales that phonons are emitted  \cite{Barenghi20082195,kobayashi:045603}.
Just as the dissipation wave number $k_\text{diss}=\mathcal E^{\frac{1}{4}}\, \nu^{-\frac{3}{4}}$ cuts off the Kolmogorov ernergy cascade in classical turbulence, the Kelvin wave cascade in quantum turbulence is truncated at the wave numbers where sound emission breaks down.  Thus a power-law spectrum is anticipated in the Kelvin wave cascade wave number range in the incompressible kinetic energy spectrum
\(
{E(k)\approx}\, k^{-\alpha},
\)
for $k\gg\xi$, where the exponent  $\alpha$ is yet to be determined.  There has been considerable effort to devise theories  and related numerical methods \cite{Barenghi20082195,kobayashi:045603,Collection1}
that yield the precise value of this exponent as well as the behavior of the incompressible kinetic energy spectrum in the transition region between the Kolmogorov and the Kelvin wave cascade spectra. 

{\it Gross-Pitaevskii equation.}--- At sufficiently low temperatures, the ground state wave function, $\varphi$, of a BEC can be described by the (normalized) GP equation
\begin{equation}
\label{Gross_Pitaevskii_equation_form1}
i \partial_t \varphi = - \nabla^2 \varphi +a (g| \varphi |^2-1 )\varphi.
\end{equation} 
We have introduced two parameters in (\ref{Gross_Pitaevskii_equation_form1}) that are useful in our numerical simulations:  $\it a$ is simply a spatial rescaling parameter to enhance the grid resolution of the vortex core and $\it g$ is a measure of the strength of the nonlinear coupling term in GP.  Unlike the Navier-Stokes equation for classical turbulence, the GP equation is a Hamiltonian system:  the total energy $E_\text{\sc tot} = const.$  It is well known 
\cite{nore:2644}
that the Madelung transformation $\varphi =\sqrt{\rho} \,e^{ i \theta}$ on the GP equation results in compressible inviscid fluid equations for the density $\rho=|\varphi|^2$ and velocity  $\mathbf{v} =2 \nabla \theta$, with the appearance of quantum pressure terms in the momentum and energy equations.  The $E_\text{\sc tot}$ can be split into an incompressible and compressible kinetic energies, an internal energy and a quantum energy:
\begin{equation}
E_\text{\sc tot} = E_\text{\tiny kin}^\text{\tiny comp}(t) + E_\text{\tiny kin}^\text{\tiny incomp}(t) + E_\text{\tiny int}(t) + E_\text{\tiny qu}(t) = const.
\end{equation} 

Typically, the GP equation has been solved numerically by the split Fourier time method \cite{kasamatsu:063616}. 
In the seminal work of Nore {\it et. al}. \cite{PhysRevLett.78.3896}, 
the GP equation was solved on a $512^3$ grid and their incompressible kinetic energy spectra while  not incompatible with the Kolmogorov $k^{-\frac{5}{3}}$ scaling did not unequivocally prove the $k^{-\frac{5}{3}}$ scaling.  
Barenghi \cite{Barenghi20082195} and
Kobayashi {\it et. al} \cite{kobayashi:045603} 
attributed this to the presence of the Kelvin wave cascade.  The Tsubota group [1] introduced a wave number dependent dissipative term into their simulations to damp out wave numbers on the order of the vortex core.  While this  suppresses the Kelvin wave cascade on the quantum turbulence, it also leads to a time decay in both the total number and total energy $E_\text{\sc tot}$.  To circumvent the decay in the total number, they add a time-varying chemical potential  in the GP equation although the total energy still decays. Most of their simulations were restricted to a $512^3$ grid and did not yield a convincing incompressible kinetic energy spectrum of $k^{-\frac{5}{3}}$ for this augmented GP equation.  These earlier simulations could not (nor did not want) to resolve the Kelvin wave cascade regime and its spectral power.

In this Letter we will, for the first time (to our knowledge), perform very high grid resolution runs (up to $5760^3$) of the Hamiltonian GP equation to resolve the quantum Kelvin wave cascade.  We use a novel unitary quantum algorithm whose solution clearly exhibits three power law regions for $E_\text{\tiny kin}^\text{\tiny incomp}(k)$: for small $\it k$ the Kolmogorov $k^{-\frac{5}{3}}$ spectrum while for high $\it k$ a Kevlin wave spectrum of $k^{-3}$.  A transitional power law on the order of $k^{-6}$ to $k^{-7}$ joins these two spectral regions.  For very large k, the phonon emissions abruptly cut off the Kelvin wave power law spectrum.

{\it Unitary quantum lattice gas algorithm.}---To recover the GP equation in 3+1 dimensions we will need only specify 2 complex numbers per point $\bm{x}$ on a cubic lattice and consider
 \begin{equation}
\label{psi_2_component_form}
\psi(\bm{x},t) = \begin{pmatrix}
\alpha(\bm{x},t)   \\
 \beta(\bm{x},t)  
\end{pmatrix}
\end{equation}
where in a quantum computer $\alpha$ and $\beta$ would be the ``excited'' state amplitudes of two qubits , respectively, at $\bm{x}$.  At each point, these excited state probability amplitudes are entangled by a collision operator of the form
 \begin{equation}
\label{collide_operator}
 C= e^{i   \frac{\pi}{4} \sigma_x (1-\sigma_x)},
  \end{equation}
where the Pauli spin matrices are
\begin{equation}
\sigma_x =\begin{pmatrix}
   0   & 1   \\
   1   &  0
\end{pmatrix}
\qquad 
\sigma_y =\begin{pmatrix}
   0   & -i   \\
   i   &  0
\end{pmatrix}
\qquad 
\sigma_z =\begin{pmatrix}
   1   & 0   \\
   0   &  -1
\end{pmatrix}.
\end{equation}
With $n=\frac{1}{2}(1-\sigma_z)$ and $\bar n=\frac{1}{2}(1+\sigma_z)$, the local qubit entanglement is then spread throughout the lattice by streaming operators 
\begin{equation}
\label{stream_operators}
S_{\Delta\bm{x}, 0} = n + e^{\Delta \bm{x} \partial_{\bm{x}}}\,\bar n,
\qquad
 S_{\Delta\bm{x}, 1} =\bar n + e^{\Delta \bm{x} \partial_{\bm{x}}} \,n,
 \end{equation}
unitarily shifting the components of $\psi$ along $\pm\Delta\bm{x}$, respectively.  
 In particular, let us consider the evolution operator for the $\gamma$th component of $\psi$.  
Our quantum algorithm interleave the noncommuting collide and stream operators, {\it i.e.} $[ S_{\Delta \bm{x},\gamma},C]\ne0$,
\begin{equation}
\label{interleaved_operator}
I_{x\gamma} =  S_{-\Delta \bm{x},\gamma}  C^\dagger  S_{\Delta \bm{x},\gamma}  C
\end{equation}
where $C^\dagger$ is the adjoint of C and $\gamma$ is either 0 or 1 corresponding to streaming either the $\alpha$ or $\beta$ component of $\psi$ in (\ref{psi_2_component_form}).  Since $|\Delta \bm{x}|$ is small, (\ref{interleaved_operator}) is close to unity.

Consider the following evolution operator for the $\gamma$ component of $\psi$
 \begin{equation}
\label{basic_typeII_quantum_algorithm}
  U_ \gamma[\Omega(\bm{x})]= I_{x\gamma}^2 I_{y \gamma}^2 I_{z \gamma}^2 e^{-i  \varepsilon^2\Omega(\bm{x})} ,
\end{equation}
 where $\varepsilon$ is a small perturbation parameter and $\Omega$ will be specified later.  With this evolution operator, the time advancement of the state $\psi$ is given by
 \begin{equation}
\label{basic_unitary_evolution_equation}
\psi(\bm{x}, t+\Delta t) =  U_ \gamma[\Omega] \,\psi(\bm{x}, t).
\end{equation}

After considerable algebra, it can be shown that for the particular unitary collide-stream operators in (\ref{collide_operator}) and (\ref{stream_operators}), one obtains the following quantum lattice gas equation on expanding in $\varepsilon$
\begin{equation}
\label{quantum_lattice_gas_equation_spinor_form}
\begin{split}
\psi(\bm{x}, t+\Delta t) = 
\psi(\bm{x}, t)
-i \varepsilon^2 \left[
-\frac{1}{2}\sigma_x\nabla^2
+\Omega 
\right]
\psi(\bm{x}, t)
\\
+
\frac{(-1)^\gamma\varepsilon^3}{4}(\sigma_y+\sigma_z)\nabla^3
\psi(\bm{x}, t)
+
{\cal O}(\varepsilon^4),
\end{split}
\end{equation}
 where $\gamma=0$ or $1$.
Since the order $\varepsilon^3$ term in (\ref{quantum_lattice_gas_equation_spinor_form}) changes sign with $\gamma$, one can eliminate this term by using a symmetrized evolution operator 
\begin{equation}
\label{symmetrized_evolution}
U[\Omega] = U_{1}\left[\frac{\Omega}{2}\right]U_{0}\left[\frac{\Omega}{2}\right].
\end{equation}
Under diffusion ordering,  in the scaling limit $\left[
\psi(\bm{x}, t+\Delta t) - 
\psi(\bm{x}, t)
\right]\rightarrow \varepsilon^2\partial_t\psi(\bm{x}, t)$, 
the quantum map $\psi(\bm{x},t+\Delta t) =  U[\Omega(\bm{x})] \,\psi(\bm{x}, t)$
 leads to a representation of the two-component parabolic equation
\begin{equation}
i \partial_t \psi = 
\left[
-\frac{1}{2}\sigma_x\nabla^2
+\Omega 
\right]
\psi(\bm{x}, t)
+
{\cal O}(\varepsilon^2),
\end{equation}
where we still have not specified the operator $\Omega$.  To recover the GP equation (\ref{Gross_Pitaevskii_equation_form1}), one simply rescales the spatial grid $\nabla \rightarrow a^{-1} \nabla$, contracts the 2-component field $\psi$ to the (scalar) BEC wave function $\varphi$
 \begin{equation}
\varphi = (1,1) \cdot \psi  = \alpha + \beta
\end{equation}
and chooses $\Omega = g |\varphi|^2-1$ :
\begin{equation}
\label{Gross_Pitaevskii_equation}
i \partial_t \varphi = - \nabla^2 \varphi +a (g| \varphi |^2 -1)\varphi\,
+
{\cal O}(\varepsilon^2).
\end{equation} 

Several comments are in order here.  $\bm{1}$ Our unitary algorithm totally respects the Hamiltonian nature of the GP equation.  No artifical numerical dissipation is introduced by our mesoscopically reversible algorithm. $\bm{2}$  While the quantum lattice algorithm is ostensibly second order accurate, it turns out that the actual accuracy of our method approaches pseudo-spectral accuracies.  This enhanced accuracy has also been noted in a somewhat related mesoscopic algorithmic cousin---the lattice Boltzmann scheme \cite{PhysRevE.68.025103,PhysRevE.75.036712,succi-2001}.  It can be traced to the imposition of detailed-balanced collisions in the mesoscopic collision routine and emergent Onsager reciprocity.  $\bm{3}$ This basic quantum lattice gas code with interleaved unitary collide-stream sequence and operator $\Omega$ has been benchmarked against exact collisions of scalar and of vector soliton solutions of the 1D  and 2D Nonlinear Schrodinger equation \cite{vahala-yepez-pt04,yepez-vahala-qip-05} as well as that for the Korteweg de Vries (KdV) solitons \cite{vahala-yepez-pla03}.  $\bm{4}$ Since our quantum algorithm consists of purely local collisions and streaming of  information to the nearby grid points, it is ideally parallel.  In fact, we have seen no saturation in performance up to the maximum number of cores available to us: 12,288 cores on the CRAY XT-5 and 131,072 cores on the IGM Blue Gene/P. $\bm{5}$  While not stressed here, our quantum lattice-gas algorithms can also be run on quantum computers when they become available;  two qubits are used to represent the $\psi$ field at a point instead of just two complex amplitudes in (\ref{psi_2_component_form}) and the collision operator (\ref{collide_operator}) is represented by a 2-qubit $\sqrt{\text{\sc swap}}$ quantum-logic gate that creates pairwise entanglement between the on-site qubits.  This entanglement spreads through the qubit system by (\ref{stream_operators}).

{\it Quantum turbulence simulations.}---Most of our simulations had as  initial conditions a set of 12 straight line vortices consisting of three groups of 4 vortices, with the group axes in the $x$, $y$ and $z$ directions.  Because the space is periodic, these lines are topologically loops. The groupings by 4 was to ensure periodicity.  Asymptotically, one can determine the form of a straight line   vortex with unit winding number using a Pad\'e approximate to the steady state solution of (\ref{Gross_Pitaevskii_equation}), following Berloff \cite{berloff-jpamg-2004}.  In polar coordinates $(r,\phi, z$), one such straight line $z$-vortex (centered at the origin) is
\begin{equation}
\label{Pade_approximant}
\varphi (r) = e^{i \phi}\,\sqrt{\frac{11 a\, r^2 (12 + a \,r^2)}{g\,[384+ a\, r^2 (128 + 11 a \,r^2)]}}  ,
\end{equation}
asymptotically.  $| \varphi | \to 1/\sqrt {g}$ as r $ \to \infty$, and $| \varphi | \sim r \sqrt {a/g} $ as r $\to 0$.  A typical isolated core radius scales as $\xi \approx (8/a)^{\frac{1}{2}}$. Two or more perpendicularly oriented line vortices are unstable and vortex entanglement ensues from such  initial conditions. 
\begin{figure}[htbp!]
\includegraphics[width=3.25in,viewport=10 190 410 300]{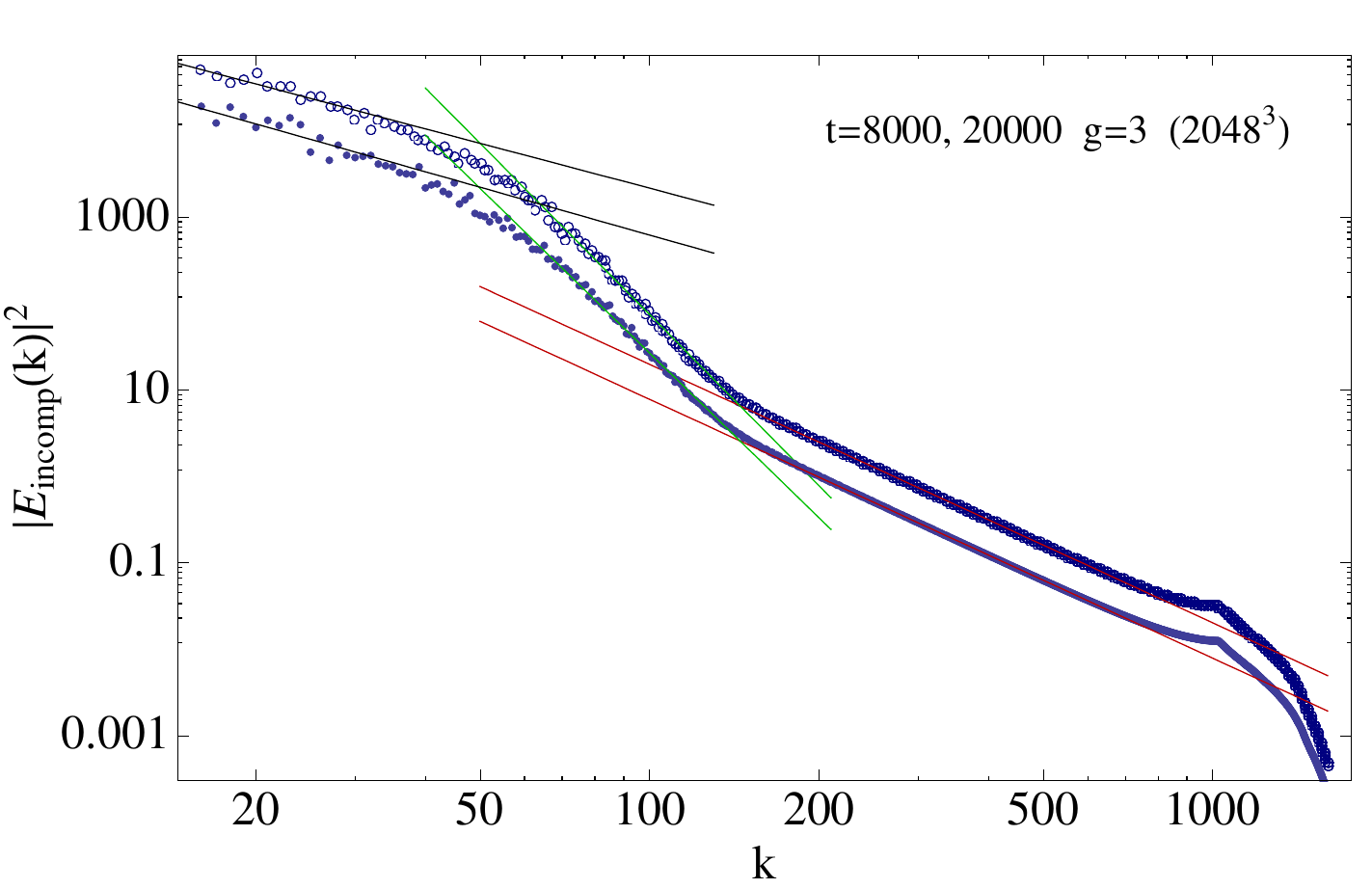}
\includegraphics[width=2.0in,viewport=40 20 350 195]{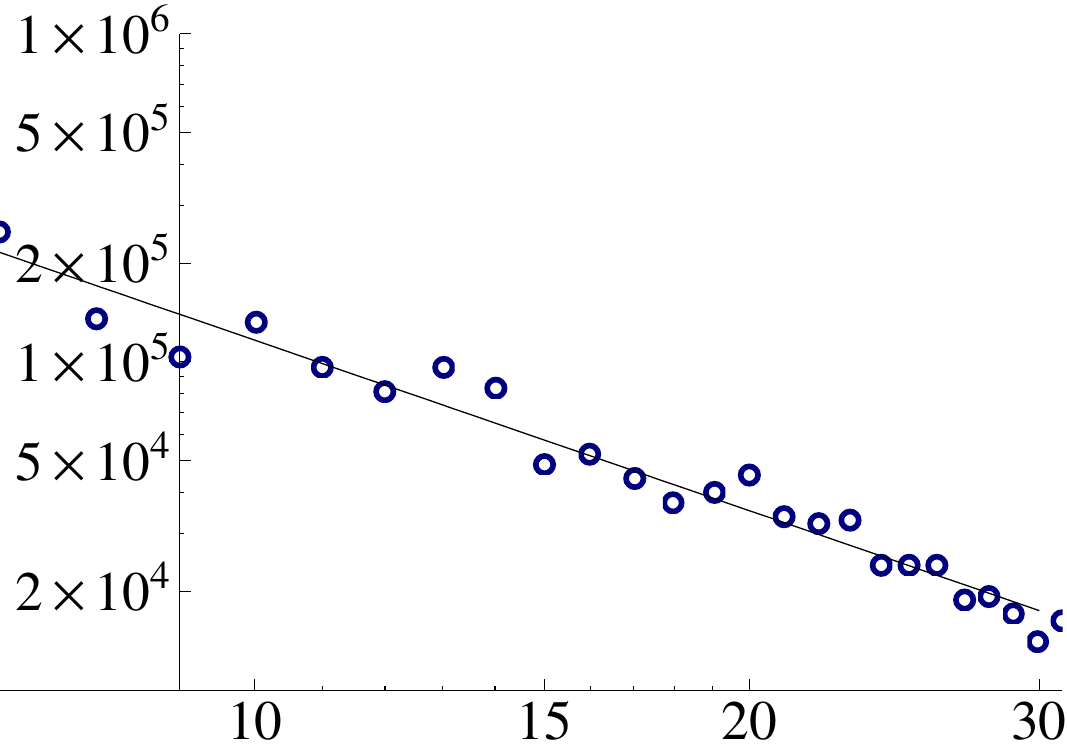}
\includegraphics[width=3.25in,viewport=10 190 410 300]{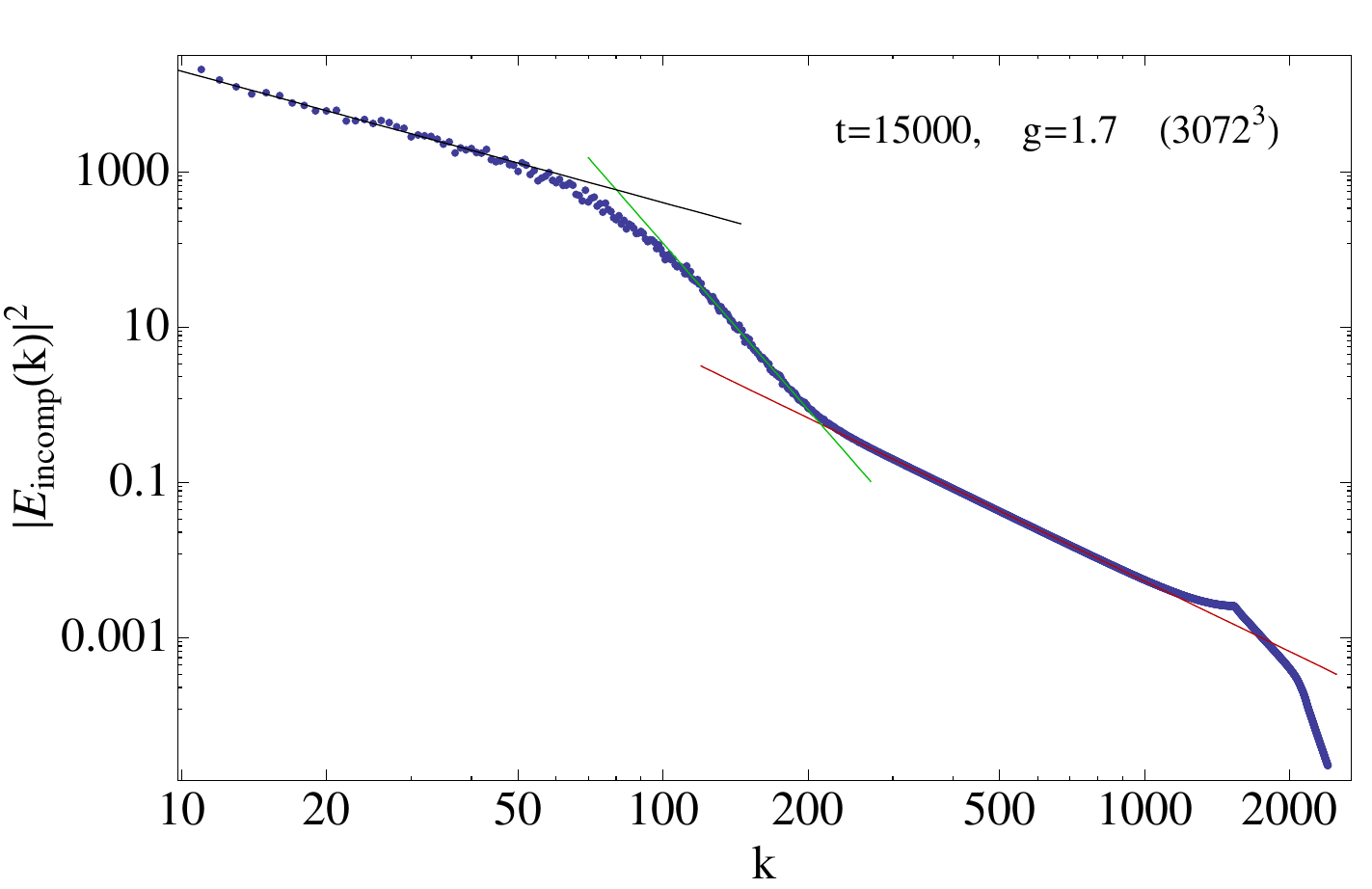}
\includegraphics[width=2.0in,viewport=40 20 350 195]{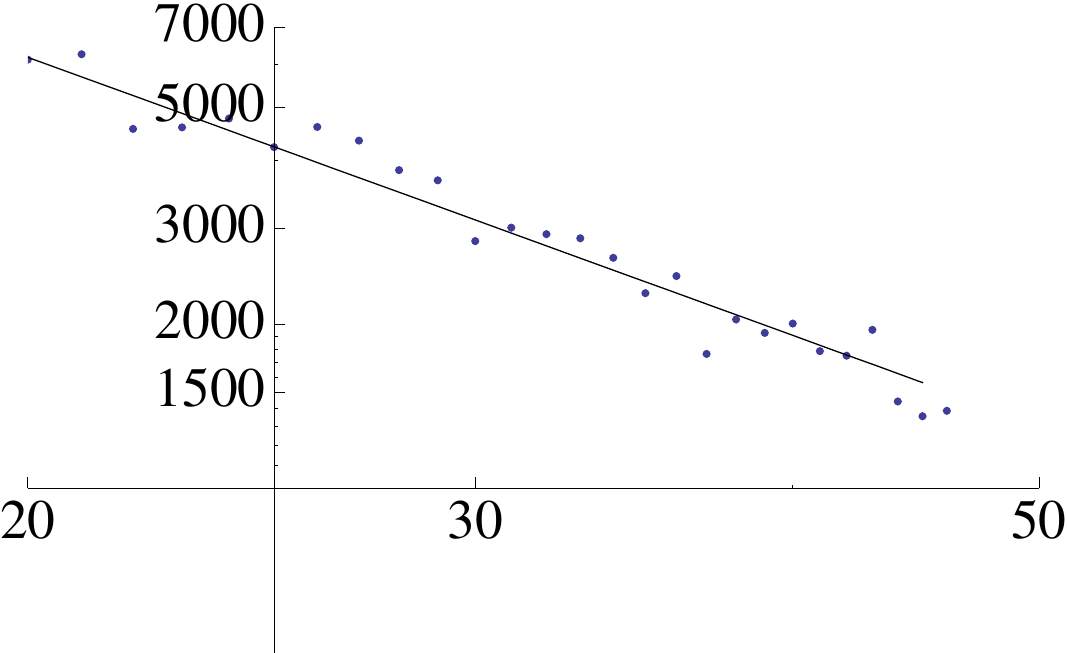}
\includegraphics[width=3.25in,viewport=10 190 410 300]{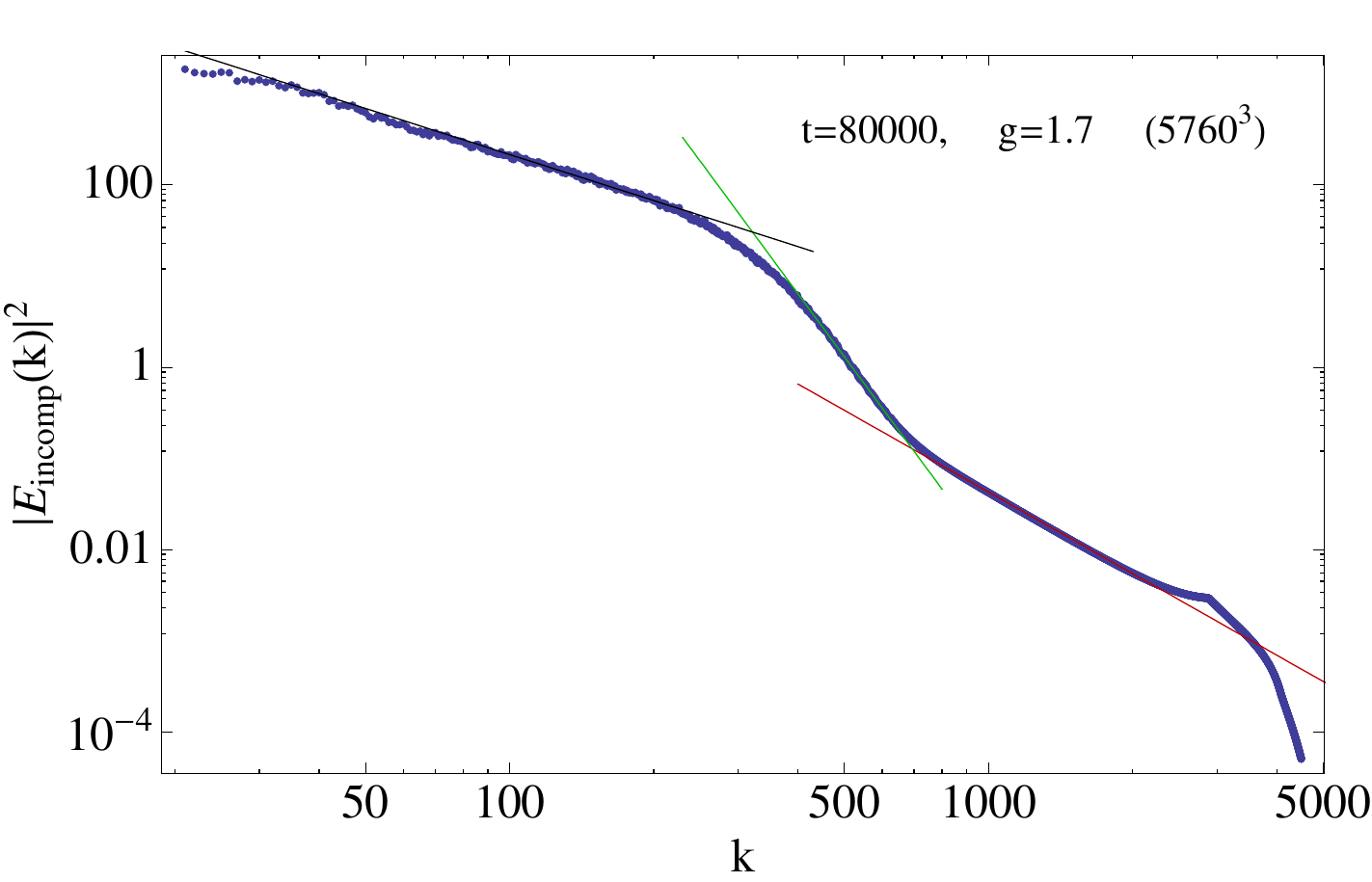}
\includegraphics[width=2.0in,viewport=40 20 350 195]{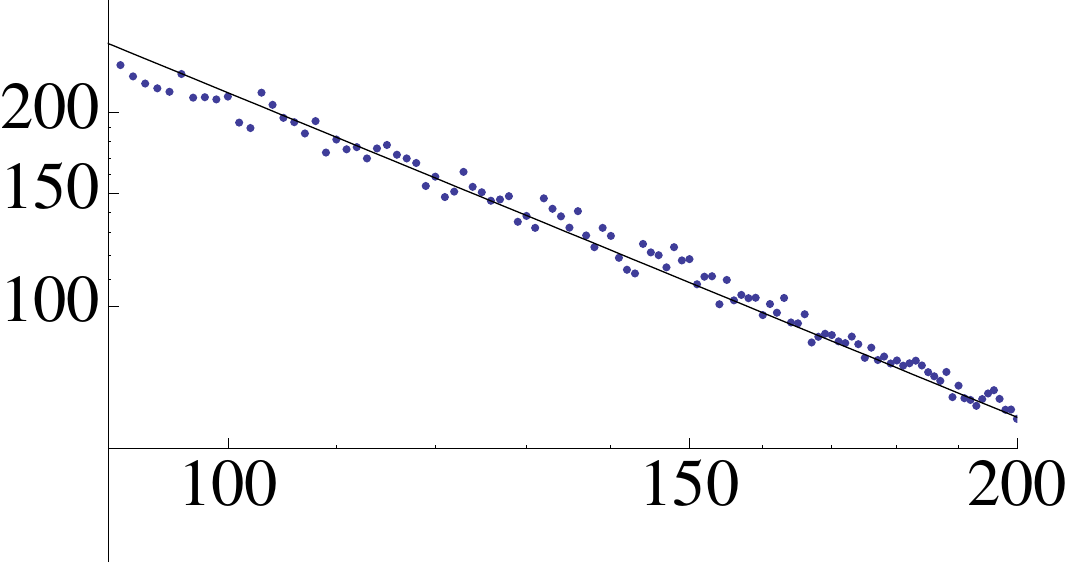}
\vspace{0.25in}
\caption{\label{spectra} \footnotesize 
The incompressible kinetic energy spectra for a periodic 12-vortex set with $a=0.02$, and an initial core radius $ \xi \approx 15\pm 5$.  The linear regression fits for power-law $k^{-\alpha}$ yield $\alpha's$ given in Table I.   There are 3 distinct spectral regions: (a) $k^{-\frac{5}{3}}$ Kolmogorov energy cascade for small $k$, (b)  steep semi-classical transition region for intermediate $k$, and (c) $k^{-3}$ Kelvin wave cascade for large $k$.
The Kolmogorov cascade becomes robust for large grids, as seen by the insets.
 }
\end{figure}
%

The incompressible kinetic energy spectrum can be extracted from the conserved total energy of the system, following Nore {\it et. al} \cite{nore:2644}.  On the top of Fig.~\ref{spectra} we present such spectra from our simulation of the GP equation on a $2048^{3}$ grid for $a=0.02$ and $g=3$ at evolution time $t = 8K$ and $t=20K$ (in lattice units).  The power laws are determined by linear regression.  
Thus, within a single simulation run, we find that the incompressible kinetic energy spectrum has three distinct power law $k^{-\alpha}$ regions that range from the classical turbulent regime of Kolmogorov for ``large'' scales (much greater than the individual quantized vortex cores) to the quantum Kelvin wave cascades at the ``small'' scales (on the order of the individual quantized cores).  There is a semi-classical region adjoining the Kolmogorov and Kelvin spectra.
These three power-law regions are quite robust as shown in middle and bottom of Figs.~\ref{spectra} from simulations on larger grids:  $3072^3$ and $5760^3$ and different initial conditions.  In particular, the initial conditions for the $5760^3$ grid simulation are chosen to have very long Poincar\'e recurrence time.  Since the GP equation is Hamiltonian, Poincar\'e recurrence exists for arbitrary initial conditions.  We found Poincar\'e recurrence to occur very rapidly for these simple 12-vortex systems (because of space limitations we shall discuss these results elsewhere).  Hence to have very large Poincar\'e recurrence times we chose initial conditions of the form $\Phi =( \Pi_{i}{\varphi_{i})}^6$.  
The exponent $\alpha$ for the spectral $k^{-\alpha}$ are given in Table~\ref{tab1}, together with the range of $k$ for these regions.
 A linear regression fit for the Kolmorogov range is shown in Fig.~\ref{spectra}.  A similar fit for the Kelvin wave range is not shown since there is no point scatter about the regression line.
The sharp drop-off in the spectrum at the end of the Kelvin wave cascade is due to the emission from very short wavelength phonons.
\begingroup
\begin{table}[thbp!]
\centering \caption{\footnotesize The spectral exponent $k^{-\alpha}$ for the 3 distinct regions, as determined by linear regression.  The first row for $2048^3$ grid is at time t = 8K while the next row for $2048^3$grid is at time t = 20K} 
\label{tab1} \smallskip
\begin{tabular}
{|l|l|l|l|} 
\hline 
{\sc Grid}
& 
{\sc Kolmogorov }
& 
{\sc Semi-classical }
&
{\sc Kelvin wave}
\\
[0.25ex]
\hline 
$2048^3$ 
& 
1.73 {\tiny ($6<k<30$)} 
& 
6.59 {\tiny $(60<k<140)$}
&
2.96 {\tiny$ (250<k<600)$} 
 \\
[0.25ex]
\hline 
$2048^3$ 
&
1.84 {\tiny ($6<k<30$)} 
&
6.34 {\tiny $(60<k<140)$}
&
2.97 {\tiny $(250<k<600)$} 
\\
[0.25ex]
\hline 
$3072^3$
& 
1.69 {\tiny ($7<k<45$)} 
& 
7.11 {\tiny ($120<k<200$)}  
& 
3.01 {\tiny ($220<k<1000$)} \\
[0.25ex]
\hline
 $5760^3$ 
& 
1.68 {\tiny ($90<k<230$)} 
& 
7.12 {\tiny ($430<k<600$)} 
&
3.00 {\tiny($1000<k<1650$)} 
\\
[0.25ex]
\hline
\end{tabular}
\end{table}
 \endgroup

{\it Acknowledgements.}---We thank S. Ziegeler for help with graphics.  We used the CRAY XT-5, 12,288 cores at NAVO and 10,240 cores at ARL.  
We are grateful for help from the administrative staff of NAVO and ARL.

\end{document}